\documentclass[runningheads]{llncs}
\usepackage[T1]{fontenc}
\usepackage{graphicx}
\usepackage{booktabs}
\usepackage{subcaption}
\usepackage{hyperref}

\begin{document}
\title{DREAMS: Diverse Reactions of Engagement and Attention Mind
States Dataset}
\titlerunning{DREAMS: Diverse Reactions of Engagement and Attention Mind States}
%
\author{Monisha Singh\inst{1}\orcidID{0000-0003-0373-8169} \and
Gulshan Sharma\inst{2,1}\orcidID{0000-0002-5332-7256} \and
Ximi Hoque\inst{2}\orcidID{0000-0001-7430-6270} \and
Abhinav Dhall\inst{3,1}\orcidID{0000-0002-2230-1440}}
\authorrunning{Singh et al.}
%
\institute{Indian Institute of Technology Ropar, India \and
Kroop AI, India \and
Flinders University, Australia
}

\maketitle 
\begin{abstract}
Active attention and engagement are important in improving users' learning experiences. Engagement refers to the level of involvement and interest individuals show towards a particular task. Attention, on the other hand, refers to a state where someone is entirely focused on a particular task with conscious awareness. Engagement and attention are different but closely linked concepts and can influence each other bidirectionally~\cite{matthews2010task}. To explore the relationship between user engagement and attention, we introduce the \textit{Diverse Reactions of Engagement and Attention Mind States (DREAMS)} dataset. The dataset includes facial video recordings of 32 users in naturalistic settings watching various stimuli to evoke diverse emotions. We then analyze user engagement and attention states in these videos by framing it as a classification problem, exploring single-task, transfer learning task, and multi-task settings. In single and transfer learning task settings, separate networks are applied to predict engagement and attention states. Whereas in multi-task settings a shared network is applied, which jointly learns to predict both engagement and attention states. Moreover, we examine participants' performance on video-based questionnaires and evaluate their perceived cognitive workload. In our findings, we observe (a) better classification performance in predicting engagement states in both transfer and multi-task learning compared to single-task learning and (b) higher engagement and attention states correlate with lower cognitive load and improved task performance. The dataset and the code are publicly available and can be accessed through \url{https://sites.google.com/view/dreams-dataset/dataset}.

\end{abstract}
\section{Introduction}

Human-Computer Interaction (HCI) has witnessed remarkable advancements in recent years, transforming how users engage with interactive systems. Within this context, \textit{Engagement} and \textit{Attention} are two intertwined concepts which have emerged as key elements in the design and evaluation of interactive systems. \textit{Engagement} mainly includes three aspects: behavioral, affective, and cognitive~\cite{o2010development}. Behavioral engagement refers to observable actions, behaviors, and interactions displayed by individuals while performing a task. Affective engagement refers to the emotions and feelings that users experience during their interaction. Cognitive engagement involves mental investment exerted by users while engaging with a particular task. On the other hand, \textit{Attention} refers to a state of focused cognitive engagement and conscious awareness directed toward a specific task.

Understanding the interplay between engagement and attention is useful for the design of interactive systems, as it significantly influences user experience. Positive engagement in interactive systems refers to users' active participation, while positive attention reflects the higher degree of focus users dedicate to the system. Conversely, negative engagement, often driven by confusing interfaces or uninspiring content, may lead to user disinterest and a shift of attention away from the intended interaction. The preceding statements suggest a positive correlation between engagement and attention, implying that higher engagement levels result in heightened attention. However, it is noteworthy that individuals who behaviorally appear engaged may not necessarily be attentive as it is plausible for one's mind to wander despite outward signs of engagement. This underscores the importance of refraining from accepting the conventional notion of a correlation between engagement and attention, but rather considering an orthogonal relationship between them. Consequently, understanding the intricate dynamics of this relationship becomes crucial for enhancing various facets of user interactions.

We aim to understand how engagement and attention are related to help content creators develop more immersive content. The main contributions of the paper are summarized below:

\begin{itemize}
\item We introduce \emph{DREAMS} (Diverse Reactions of Engagement and Attention Mind States), a self-annotated engagement and attention state dataset, collected in an in-the-wild setting.

\item We designed single, transfer learning, and multi task experiments to evaluate engagement and attention, assuming that the performance of related tasks would exhibit improvement in transfer learning and multi task setup compared to single task setup.

\item We applied NASA Task Load Index(NASA-TLX) workload~\cite{hart1988development} assessment questions to identify the variations in cognitive load levels experienced by individuals while watching the diverse set of stimuli. Additionally, we explored the causal relationship between engagement, attention, and cognitive load levels.

\item We study the impact of engagement and attention on task performance by analyzing the percentage of correct responses for various engagement and attention levels.

\end{itemize}

\begin{figure}[t]
    \centering
    \begin{subfigure}[b]{0.45\textwidth}
        \includegraphics[scale = 0.125]{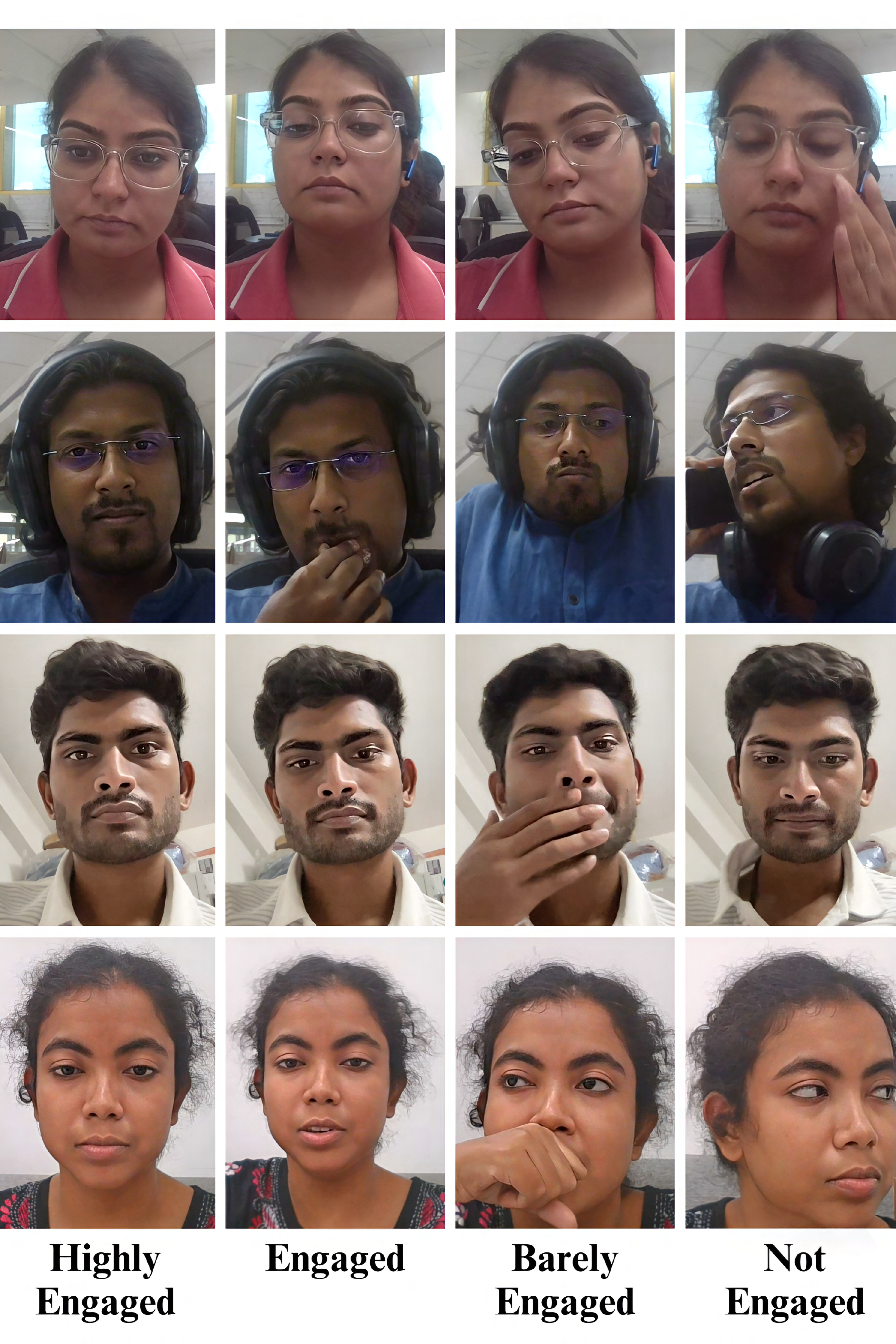}
        \label{fig:engagement_frames}
    \end{subfigure}
    \hfill
    \begin{subfigure}[b]{0.45\textwidth}
        \includegraphics[scale = 0.125]{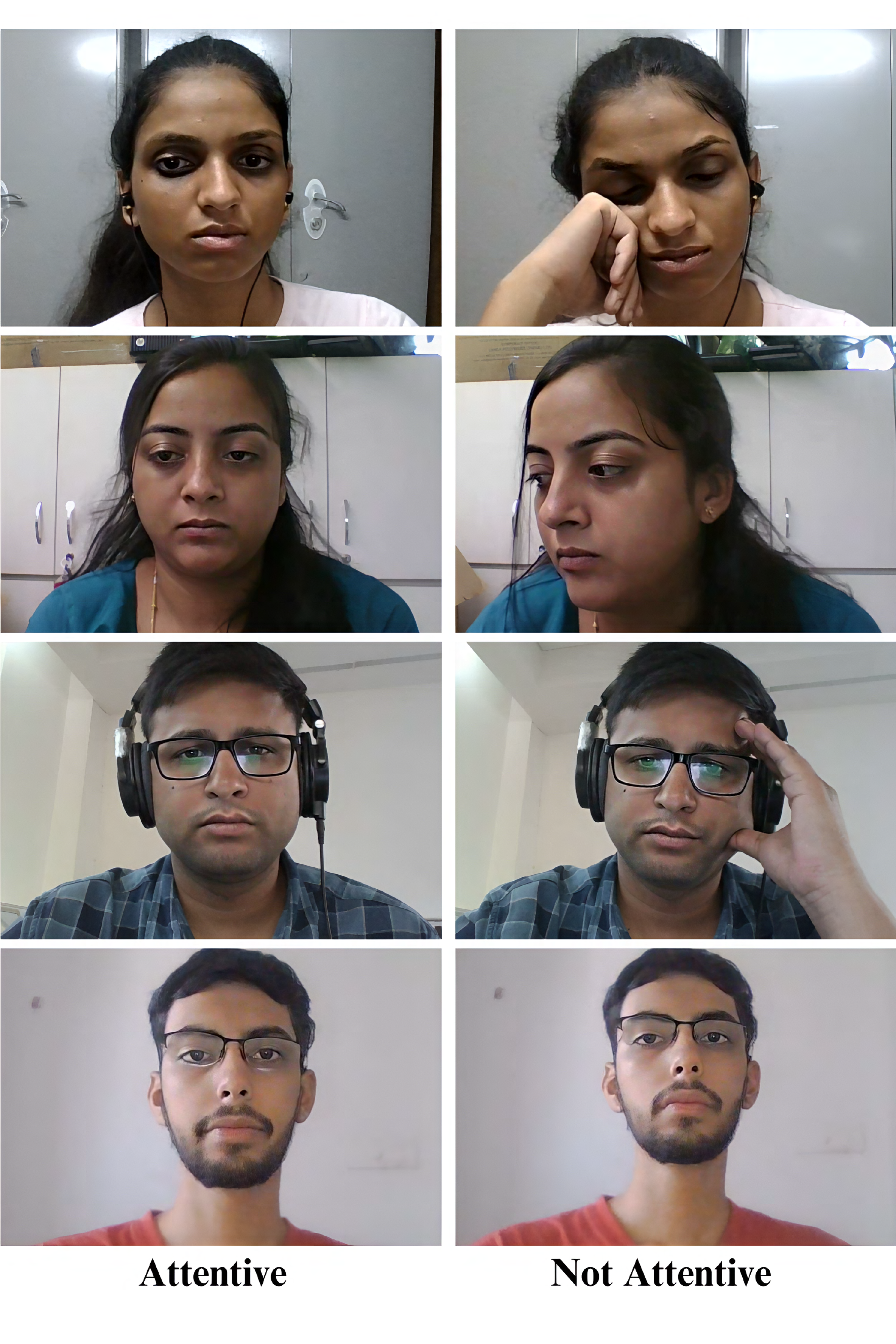}
        \label{fig:attention_frames}
    \end{subfigure}
    \caption{Engagement(left) and Attention(right) states from the \emph{DREAMS} dataset. On the left, notice that \emph{Highly Engaged} participants are glued to the screen. Subtle changes in expressions and gestures can be observed as engagement decreases gradually.
    On the right, notice that outward signs of attention are displayed by the first-three-row participants. However, the last-row participant appears to be visually attentive in both states, but his mind may have been wandering in the second instance.}
    \label{fig:frames}
\vspace{-0.3cm}
\end{figure}

\section{Background \& Related Work}

In this section, we present an overview of contributions and recent advancements in the fields of engagement and attention research.
\subsection{User Engagement}
Engagement encompasses users' observable actions (Behavioral), emotional connection (Affective), and cognitive investment~\cite{fredricks2004school}. Behavioral engagement can be assessed through qualitative observations of user behavior, such as gaze, head pose, facial expressions, etc. Various methods, including surveys, physiological measures, behavioral analysis, and neuroimaging techniques, can measure affective engagement. Multiple approaches, such as evaluating task performance, self-reporting, analyzing response times, monitoring physiological indicators, and employing neuroimaging techniques, can be applied to gauge cognitive engagement. User engagement plays a crucial role in achieving various goals, including educational success, productivity, customer satisfaction, and business growth. User engagement has been extensively studied in the context of HCI~\cite{gupta2016daisee,kaur2018prediction,10.1145/3577190.3614164}. One of the earliest efforts in this direction is by D'Mello et al.\cite{10.1504/IJLT.2009.028805}. They investigated the relationship between facial expressions, linguistic cues, and engagement detection and explained the importance of non-verbal cues in comprehending user engagement. Facial Action Coding System was used by authors in \cite{grafsgaard2013automatically,whitehill2014faces} to assess distinct emotions associated with various facial muscle movements to emphasize the relationship between particular engagement labels and facial action units. 
In \cite{whitehill2014faces}, Whitehill et al. showed that automated engagement detectors work as accurately as people using non-verbal features. To assess engagement, Booth et al. \cite{booth2017toward} evaluated the performance of an LSTM-based method. In the EmotiW 2018~\cite{10.1145/3242969.3264993} sub-challenge, \emph{Engagement Prediction in the Wild}, authors proposed using TCN \cite{thomas2018predicting} to predict user engagement in an in-the-wild setting. TCN network was applied here to show improvements over LSTM in improving baseline models. In the recent works, authors tried to understand patterns that relate to User Engagement by demonstrating and focusing on interpretability and having a simpler SVM model to train for predicting engagement\cite{stappen2022estimation}.

\subsection{User Attention}
Attention can be described as the mental state in which an individual directs their entire focus and conscious awareness toward a particular task or stimulus. Attention is vital for selectively processing information, facilitating cognitive functions, and enabling efficient task performance. Attention can be measured through various methods such as reaction time tasks, eye-tracking technology, neuroimaging techniques like Electroencephalography or Functional Magnetic Resonance Imaging, and behavioral observation. Attention, a crucial process, is one of the main focus of HCI research. Eye-tracking experiments have greatly added to the study of visual attention patterns. A significant contribution was made by Buscher et al.~\cite{10.1145/1518701.1518705} in which they investigated the dynamics of visual attention while browsing the web. Their research provided an understanding of the hierarchy of visual saliency and user preferences by demonstrating how consumers distribute their attention across various web page elements. The work by D'Mello et al.~\cite{10.1016/j.ijhcs.2012.01.004} stands out in the field of attention-aware, intelligent tutoring systems. In their study, they introduce a pioneering approach that harnesses gaze data to enhance the effectiveness of tutoring interactions. The Gaze Tutor system employs eye-tracking technology to track students' gaze patterns and alter instructional content based on their visual attention. Their work demonstrates the potential of attention-aware technology in building more responsive and personalized educational experiences by smoothly incorporating gaze tracking into the tutoring process. 

In their study, Hutt et al.~\cite{10.1145/3411764.3445269} offer a compelling investigation into the use of gaze-based attention-aware technologies to combat students' tendency to daydream in classroom environments. The authors suggest a novel strategy that uses gaze monitoring to identify instances of mind wandering and then implements targeted interventions to draw students' attention back. This work provides important insights into the practical implications and efficacy of such interventions inside classroom situations by expanding the deployment of gaze-based attention-aware technologies outside controlled laboratory settings. Recent work in attention(mind wandering) detection by Lee et al.~\cite{lee2022predicting} involved the use of facial action units extracted from videos of students participating in online learning. The authors emphasized that their proposed feature-based model outperformed random and gaze-only baselines and suggested using landmark-based features over gaze features.

Researchers from various fields have been intrigued by the complex interplay between engagement and attention, which has provided light on the dynamic nature of human cognition and interaction. In his study, Heath~\cite{heath2007we} investigates how attention and engagement mutually influence each other, contributing to the overall impact of advertisements on viewers. His research reveals a strange pattern. A higher level of emotional content in the advertising correlates with lower levels of attentiveness. This discovery casts doubt that engagement and attention have a simple causal relationship. Instead, the experiment demonstrates that engagement and attention can function separately and are not necessarily associated. The study by Leiker et al.~\cite{leiker2016relationship} used neurophysiological measures of attention to investigate the relationship between engagement and attention in motion-controlled video games. This study found that engagement elicits increased information processing, which reduces attentional reserve. These initiatives underscores the importance of considering both engagement and attention in designing effective educational interventions and interactive systems.

\section{DREAMS Dataset}
\label{section:stimuli_and_experiment_flow}
\subsection{Ethics}
We obtain user consent for the collection of their data and meta-information, ensuring transparency and respect for individual autonomy. Also, all stimuli used in the study are sourced either from publicly available datasets or videos having Creative Commons licenses. By doing this, we protect user's rights and promote transparency in our research processes.

\subsection{Stimuli}
We use three stimuli from the EngageNet~\cite{10.1145/3577190.3614164} dataset, (i) \emph{Schrödinger's cat: A thought experiment in quantum mechanics} (ii) \emph{What is cryptocurrency?} (iii) \emph{Where did English come from?}. 
The first and second stimuli feature a digital teacher avatar (refer Figure 2). These stimuli provide instruction on quantum mechanics and cryptocurrency, respectively.
While the third stimulus offers insights into the evolution of the English language. The avatar based videos are generated using the Artiste platform from Kroop AI. Additionally, we include two humorous videos with Creative Commons licenses to diversify the stimuli set. The educational stimuli typically have an average duration of 5 minutes, whereas the humorous stimuli are shorter, lasting approximately for 2 minutes. The background sources of the stimuli can be found here\footnote{Video Source: \href{https://www.youtube.com/watch?v=UjaAxUO6-Uw}{Quantum Mechanics}, \href{https://www.youtube.com/watch?v=1YyAzVmP9xQ}{Cryptocurrency}, \href{https://www.youtube.com/watch?v=YEaSxhcns7Y}{English Language},
\href{https://www.youtube.com/watch?v=ZnjJpa1LBOY}{Humorous Stimulus 1}, \href{https://www.youtube.com/watch?v=Q1z9AU7_e-g}{Humorous Stimulus 2}}.

\subsection{Data Collection Protocol}
We develop a web-based interface to collect data on user's engagement and attention states. The data is collected in an in-the-wild settings, where participants have the flexibility to record data at their preferred time and location. Participants can use a computer or laptop with a reliable internet connection and a good webcam. The experiment starts with collecting participants' demographic data, including biological sex and age. Afterwards, a pre-study questionnaire is given to gauge user personality traits. 

Following the completion of the questionnaire, the video stimuli are presented in a randomized order. As the stimuli play, automatic prompts are integrated into the web interface, asking participants to self-assess their attention and engagement states based on certain guidelines provided to them.

For attention state, the participants could choose from the following options: \emph{Focused/Attentive} (thinking about the stimulus), \emph{Not Focused/Not Attentive} (mind-wandered, doing or thinking something unrelated to the lecture), or \emph{Skip} (participant is indecisive about his/her state). 

For engagement, participants could choose from the following engagement levels: \emph{Highly Engaged} (participant is attentive and glued to the screen), \emph{Engaged} (participant is interested in the content, and appeared to like it), \emph{Barely Engaged} (participant is minimally attentive, fidgeted restlessly in the chair or hardly opened his or her eyes), and \emph{Not Engaged} (participant are disengaged, frequently glanced away from the screen and are disinterested). 

After completion of a stimuli, participants are required to fill out a brief questionnaire, which includes stimulus content based questions, engagement-related questions and cognitive load assessments using NASA-TLX. The typical duration of a user session varies from 25 to 30 minutes, depending on the questionnaires' response time of participants.

\subsection{Participants} The study involved 32 college students, comprising 13 females, with age ranging from 21 to 38 years. All participants are proficient in English with educational backgrounds in science and engineering.

\begin{figure}[t]
\centering
\includegraphics[scale=0.48]{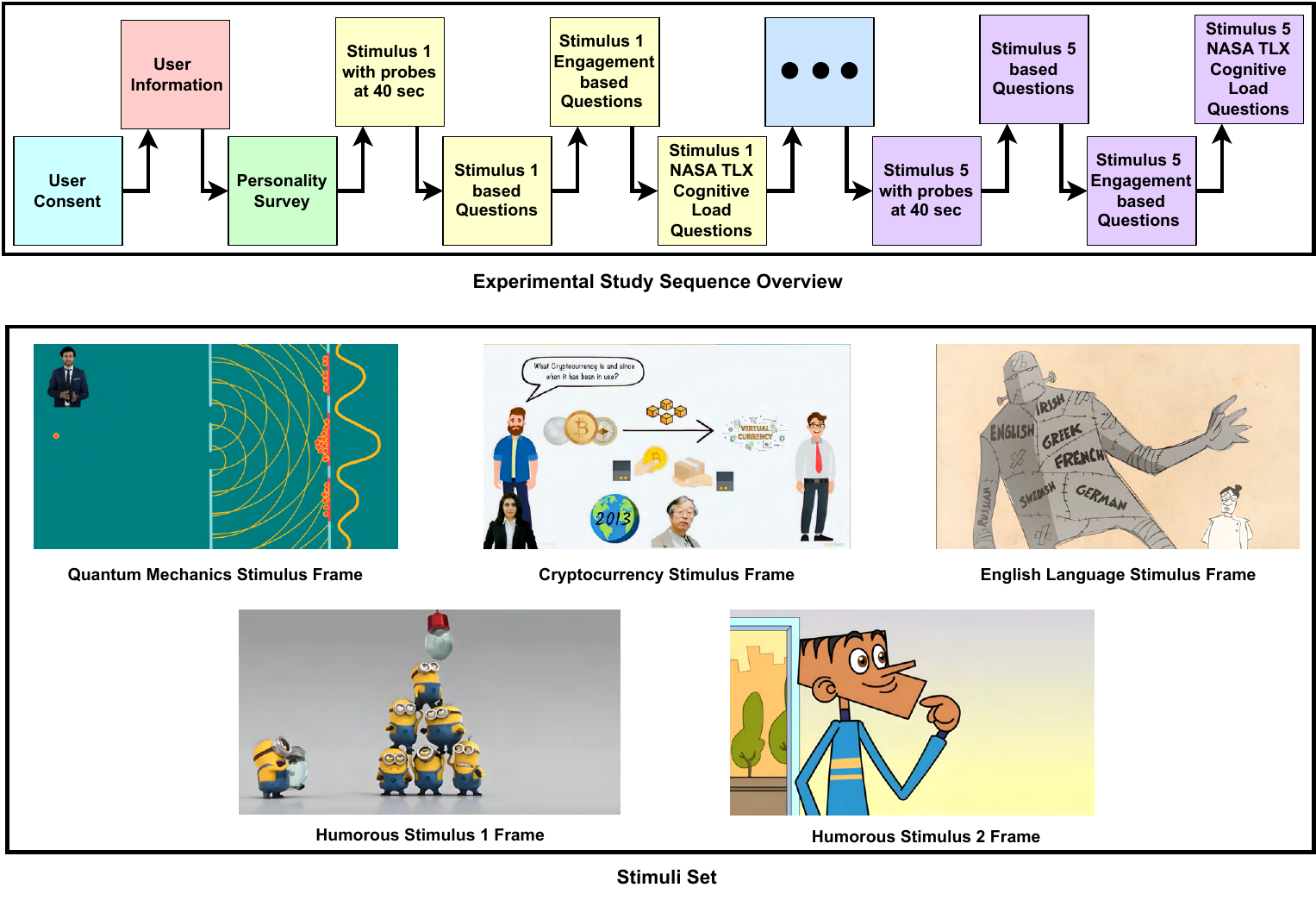}
\caption{(Top) Sequence Flow of Experimental Study, and (Bottom) Frames from Stimuli Set used in the study. Notice the digital avatars in upper left corner of the Quantum Mechanics Stimulus Frame, and in lower left corner of the Cryptocurrency Stimulus Frame. For details, refer to section~\ref{section:stimuli_and_experiment_flow}}. 

\label{fig:experimental_flow_stimuli_set}
\vspace{-0.6cm}
\end{figure}
\subsection{Data Collection \& Annotations}

During the experiment, we collect the following information from each participant: (i) responses to the personality questionnaire, (ii) video recordings of participants as they watch the stimuli, (iii) self-reported engagement and attention states provided by the participants, (iv) responses to questionnaires regarding the content of the viewed stimuli, and (v) responses to questions assessing behavioral, cognitive, and affective engagement, followed by NASA-TLX responses.

Automatic prompts are generated every 40 seconds, asking participants to self-identify their attention and engagement states. We chose a 40-second interval based on the typical time it takes for attention to shift~\cite{klinger1978modes} and the findings of an in-lab pilot study conducted by Lee et al.~\cite{lee2022predicting}. We collected 832 videos of 40 seconds duration, out of which 781 videos were successfully processed. The resolution of the recorded videos were at least 640x480 pixels.

As a pre-processing step, we combine the \emph{Not Focused/Not Attentive} and \emph{Skip} labels of attention into a single category called \emph{Not Attentive}. The rationale behind this merging was the participants’ lack of attention to the task, which prevented them from making a decision.

Table~\ref{tab:self_labels_distribution} illustrates the distribution of self-labeled instances of Engagement and Attention. To quantify the relationship between these self-labels, we applied \textit{Cramér's V} correlation coefficient. The coefficient for the self-labeled data is 0.7289, signifying a strong association between engagement and attention.

\subsection{Data Split}
We split the data into subject independent sets, with 25 subjects in \emph{Train} and 7 subjects in \emph{Test} set. There are 600 videos in the \emph{Train} set whereas 181 videos in the \emph{Test} set, amounting to a total of approximately 9 hours of data.

\begin{table}[htbp]
\caption{Distribution of self labelled instances under various Engagement and Attention states.}
\begin{center}

\begin{tabular}{lcc}
& Attentive & Not-Attentive\\
\midrule
Highly Engaged & 260 & 2\\
Engaged & 220 & 59 \\
Barely Engaged & 32 & 133 \\
Not Engaged & 8 & 67 \\
\end{tabular}%
\end{center}
\label{tab:self_labels_distribution}
\vspace{-0.5cm}
\end{table}

\section{Identifying Relationship between Attention \& Engagement}
In psychology studies, attention and engagement are closely connected concepts. Multiple studies~\cite{reeve2011agency,fredricks2012measurement,skinner2009motivational} have identified attention as an important factor in determining behavioral engagement. However, authors in~\cite{leiker2016relationship} observe that higher engagement is associated with low \textit{eP3a} levels. \textit{eP3a} is a brain wave component typically observed in response to stimuli, where higher levels are usually associated with increased attention. 
These studies suggests that the relationship between engagement and attention is complex and sometimes contradictory. Hence it is important to understand this relationship prior to devising effective learning strategies.

To understand this relationship, we apply supervised learning based approach, where we explore single-task learning, transfer learning, and multi-task learning. 

\begin{itemize}
    \item In single-task learning, we apply separate neural networks to independently learn about attention and engagement states. The idea behind this approach is to benchmark the performance of each network in identifying attention and engagement states.

    \item In transfer learning, we first pretrain the network on engagement or attention task, and later fine-tune on the related counterpart task. The idea behind this approach is to transfer the learned feature representations from one task to another to improve performance.
    
    \item In multi-task learning, we train a unified network which jointly learns to identify attention and engagement states. The idea behind this approach is to learn shared representations, allowing the network to capture underlying correlations and dependencies between attention and engagement.
\end{itemize}

\begin{figure}[htbp]
\centering
\includegraphics[scale=0.55]{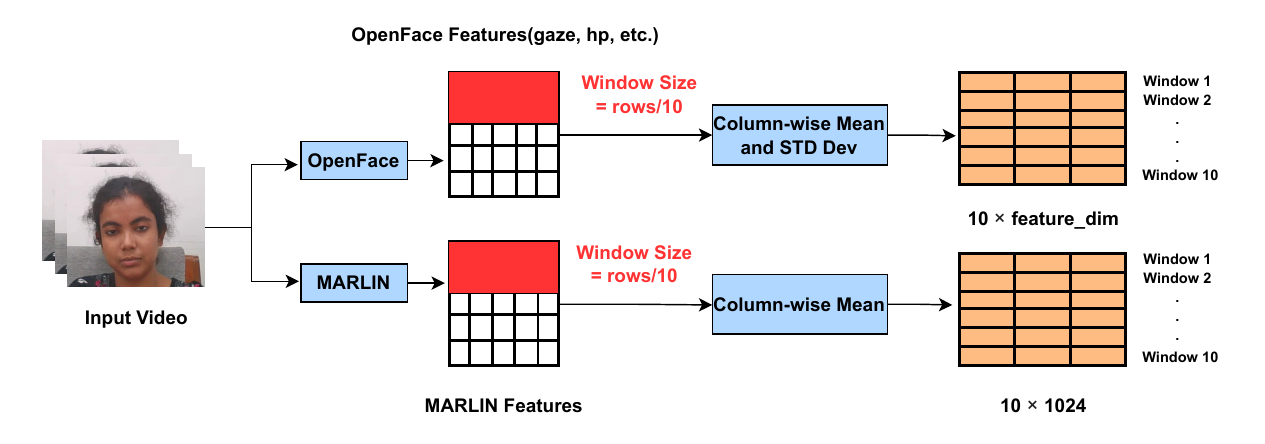}
\caption{Feature extraction and statistical feature aggregation. For details, refer to section~\ref{sect:feat}.}
\label{fig:feature_extraction}
\end{figure}

\begin{figure}[t]
\centering
\includegraphics[scale=0.5]{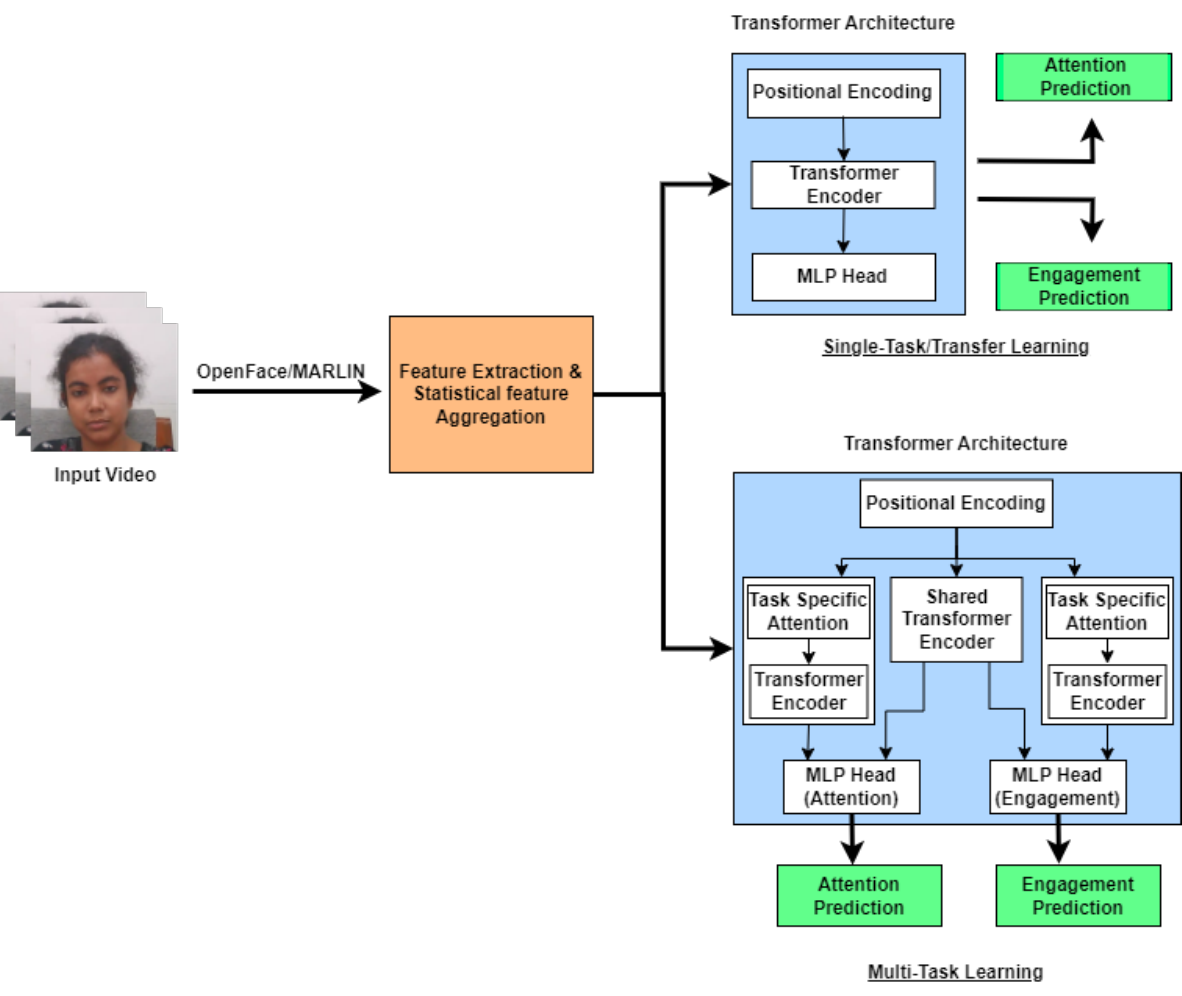}
\caption{Single-Task/Transfer Learning and Multi-Task Learning. For details, refer to section~\ref{section:models}.}
\label{fig:single_task_multi_task}
\end{figure}

\subsection{Feature Extraction}
\label{sect:feat}

We extract facial features such as eye-gaze, facial landmarks, head pose, facial action units, and Point Distribution Model (PDM) by applying OpenFace~\cite{10.1109/FG.2018.00019} framework on user videos. Additionally, we extract spatiotemporal representation of the face by applying MARLIN~\cite{cai2022marlin} framework.

After extracting facial features and
spatiotemporal representations using OpenFace and MARLIN, respectively, we partition the extracted row-wise data into 10 evenly distributed segments. Then, we compute the mean and standard deviation for each segment and combine them into a single vector. This vector statistically captures the changes in facial features over time in the video. The overall process of feature extraction is depicted in Figure~\ref{fig:feature_extraction}.

\subsection{Experiments}
\label{section:models}
To understand the relationship between engagement and attention, we apply classification methods based on Transformer~\cite{10.5555/3295222.3295349} architecture. Transformers can learn temporal and spatial relationships in the data. The attention mechanism within the Transformer allows for identifying important relationships in the data and effectively managing long-range relationships. 

This section describes the supervised methods employed to uncover the relationship between engagement and attention. Figure~\ref{fig:single_task_multi_task} illustrates these classification methods. 

\subsubsection{Single Task and Transfer Learning Task} 
Single-task learning is a methodology that optimizes a single objective function using a dedicated feature representation. Transfer learning, on the other hand, leverages knowledge gained from solving one problem and applies it to a different but related problem. We conducted single-task and transfer learning experiments using a Transformer architecture incorporating positional encoding, a transformer encoder, and a Multi-Layer Perceptron (MLP) for engagement and attention prediction. Positional encoding enhances the input data while maintaining its sequential context. The order of each component inside the sequence is considered, which is crucial for tasks involving temporal or spatial interactions. 

\subsubsection{Multi-task Learning}
Multi-task learning is a way to learn multiple objective functions using a common feature representation. We achieve this by using the same time-series transformer-based model to train on the OpenFace and MARLIN features. In the multi-task network, we have a common positional encoding and a transformer encoder layer. The shared transformer encoder layer serves as a foundational feature extractor. This layer is excellent at capturing intricate dependencies and interactions among input sequences. Two different MLP, one for attention (mind wandering) and the other for engagement, handle the encoded features. The task-specific MLP heads are designed to extract task-specific features from the shared transformer encoder's output. This specialization enables the model to excel at each task's unique challenges. Over the positional encodings, we add task-specific attention layers and establish a residual connection with MLP heads. These task-specific attention layers grant our model the ability to dynamically distribute attention across the input sequence in accordance with the unique demands of each task. The network thus benefits from a common transformer encoder layer and task-specific attention from the positional embeddings. We train the model where we try to minimize engagement and attention losses jointly, using the same encoder. The network splits into two task-specific MLP where they reduce their individual losses. 

We use cross entropy loss with assigned class weights for engagement and attention prediction. We choose the best model by maximizing the sum of the weighted F1 score, weighted Precision, and weighted Recall for engagement and attention prediction. 

\begin{table}[t]
\caption{Results of Engagement prediction in Single-Task, Transfer Learning and Multi-Task setups. Here HP, AU, LMK, and PDM refers to Head Pose, Action Units, Landmarks, and Point Distribution Model respectively. P, F1, and R refer to Weighted Precision, F1, and Recall scores respectively.}
\begin{center}

\resizebox{\linewidth
}{!}{%
\begin{tabular}{l|c|c|c|c|c|c|c|c|c}
                   & \multicolumn{9}{c}{\textbf{Engagement}}
                   \\
                   \hline
                   & \multicolumn{3}{c|}{Single-Task}          & \multicolumn{3}{c}{Transfer Learning}
                   & \multicolumn{3}{|c}{Multi-Task}
                   \\
                   \hline
Features  & P & F1 & R & P & F1 & R & P & F1 & R \\
\hline
GAZE & \textbf{0.125} & \textbf{0.185} & \textbf{0.354} & \textbf{0.125} & \textbf{0.185} & \textbf{0.354} & \textbf{0.125} & \textbf{0.185} & \textbf{0.354} \\
GAZE+HP & \textbf{0.178} & 0.181 & \textbf{0.332} & 0.110 & 0.165 & 0.331 & 0.163 & \textbf{0.218} & \textbf{0.332} \\
GAZE+HP+AU & 0.219 & 0.256 & 0.320 & 0.167 & 0.216 & \textbf{0.354} & \textbf{0.295} & \textbf{0.306} & 0.332 \\
GAZE+HP+AU+LMK & 0.496 & 0.226 & 0.320 & 0.400 & \textbf{0.293} & \textbf{0.392} & \textbf{0.596} & 0.208 & 0.365 \\
GAZE+HP+AU+LMK+PDM & 0.260 & \textbf{0.305} & 0.376 & \textbf{0.606} & 0.284 & \textbf{0.398} & 0.311 & 0.299 & 0.343 \\
MARLIN & 0.267 & 0.256 & 0.298 & 0.310 & 0.275 & 0.326 & \textbf{0.382} & \textbf{0.276} & \textbf{0.354}
\\
\hline
\end{tabular}%
}
\end{center}
\label{tab:engagement_results}
\vspace{-0.6cm}
\end{table}

\subsection{Results}
The experiments were performed in single-task, transfer learning, and multi-task setups using supervised approach. We considered gaze as our base feature, a crucial indicator of attention and behavioral engagement, and incrementally added other OpenFace features to assess their contribution in predicting engagement and attention levels. For performance comparisons we took the summation of weighted precision, weighted F1, and weighted recall scores. Below is a detailed analysis of the results of the various experiments performed as part of this study. 

\subsubsection{Single-Task vs Transfer Learning} 
From the engagement results, (refer to Table~\ref{tab:engagement_results}) we can observe that for the feature combination of Gaze + HP + AU + LMK, Gaze + HP + AU + LMK + PDM, and MARLIN, there is an improvement in the performance of transfer learning(attention pre-training and engagement fine-tuning) over the single-task learning. However, in the attention results (refer to Table ~\ref{tab:attention_results}) no improvement is observed in the performance of transfer learning(engagement pre-training and attention fine-tuning) over the single-task learning for any feature combination. Based on these observations, one can infer that attention may serve as a clearer indication of engagement, but engagement may not be a reliable indicator of attention.

\begin{table}[htbp]
\caption{Results of Attention prediction in Single-Task, Transfer Learning and Multi-Task setups. Here HP, AU, LMK, and PDM refers to Head Pose, Action Units, Landmarks, and Point Distribution Model respectively. P, F1, and R refer to Weighted Precision, F1, and Recall scores respectively.}
\begin{center}

\resizebox{\linewidth
}{!}{%
\begin{tabular}{l|c|c|c|c|c|c|c|c|c}
                   & \multicolumn{9}{c}{\textbf{Attention}}
                   \\
                   \hline
                   & \multicolumn{3}{c|}{Single-Task}          & \multicolumn{3}{c}{Transfer Learning}
                   & \multicolumn{3}{|c}{Multi-Task}
                   \\
                   \hline
Features  & P & F1 & R & P & F1 & R & P & F1 & R \\
\hline
GAZE & \textbf{0.763} & \textbf{0.465} & \textbf{0.608} & 0.158 & 0.226 & 0.398 & 0.158 & 0.226 & 0.398 \\
GAZE+HP & \textbf{0.775} & 0.534 & \textbf{0.641} & 0.564 & \textbf{0.566} & 0.580 & 0.158 & 0.226 & 0.398 \\
GAZE+HP+AU & \textbf{0.703} & 0.574 & \textbf{0.652} & 0.690 & \textbf{0.579} & \textbf{0.652} & 0.158 & 0.226 & 0.398 \\
GAZE+HP+AU+LMK & \textbf{0.675} & 0.615 & \textbf{0.663} & 0.646 & \textbf{0.617} & 0.652 & 0.363 & 0.453 & 0.602 \\
GAZE+HP+AU+LMK+PDM & 0.668 & \textbf{0.637} & \textbf{0.669} & 0.641 & 0.606 & 0.646 & \textbf{0.763} & 0.465 & 0.608 \\
MARLIN & \textbf{0.643} & \textbf{0.623} & \textbf{0.619} & 0.630 & 0.622 & \textbf{0.619} & 0.621 & 0.546 & 0.553
\\
\hline
\end{tabular}%
}
\end{center} 
\label{tab:attention_results}
\vspace{-0.6cm}
\end{table}

\subsubsection{Single-Task vs Multi-Task} 
From the engagement results table~\ref{tab:engagement_results}, we can observe that for all feature combinations except Gaze, there is an improvement in performance of multi-task over the single-task learning. However, in attention results table ~\ref{tab:attention_results} no improvement was observed in performance of multi-task over single-task learning for any feature combination. The results of this experiment also suggest that the presence or absence of attention directly affects engagement, but the reverse relationship may not be as strong or direct. In other words, the model might find it easier to learn features or patterns related to engagement when attention information is available, as attention could be a contributing factor to engagement. However, the absence of engagement information may not have a substantial impact on the model's ability to predict attention.

\subsection{Engagement and Attention NASA-TLX workload Analysis}

We applied NASA-TLX assessment questions to assess the perceived workload experienced by the participants while watching the diverse set of stimuli.  
The NASA-TLX, which originally had six arbitrary subscales, rates the following aspects: (i) Mental Demand, (ii) Physical Demand, (iii) Temporal Demand, (iv) Performance, (v) Effort, and (vi) Frustration Level.

In order to calculate the average task load for this study, we concentrated on three distinct subscales: Mental Demand, Effort, and Frustration. Using a four-point scale, our assessment classified the task load as \emph{very low}, \emph{low}, \emph{high}, or \emph{very high}. This method gave insightful information about the perceived workload related to the tasks under evaluation.

The assessment of workload across the different educational video categories yielded distinct findings. Specifically:
\begin{itemize}
\item The average workload associated with the Quantum Mechanics educational video fell within the range between \emph{low} and \emph{high}. This indicates that viewers perceived a moderate level of mental demand and effort while engaging with this educational content.

\item In contrast, both the Bitcoin and English language educational videos exhibited an average workload categorized as \emph{low}. This suggests that these videos required relatively less mental demand and effort compared to the Quantum Mechanics video.

\item Notably, the category of humorous videos was found to have an exceptionally \emph{low} workload. This implies that the act of viewing humorous content involved minimal mental demand, required little effort, and was associated with a \emph{low} level of frustration.
\end{itemize}

Our observations further revealed intriguing patterns in workload perception among participants with varying levels of engagement and attention. Specifically:
\begin{itemize}
\item Participants classified as \emph{Highly Engaged} and \emph{Engaged} consistently reported lower workload scores. This phenomenon can be attributed to their elevated levels of engagement, which likely facilitated memory retention, and content recall. Consequently, these individuals encountered reduced mental demand, spent less effort, and experienced diminished frustration when responding to questions related to the viewed content.

\item Similarly, participants categorized as \emph{Attentive} exhibited workload scores that aligned with the trend observed among \emph{Highly Engaged} and \emph{Engaged} subjects. Their attentive demeanor likely contributed to a smoother cognitive process, translating to a lower perceived workload.

\item In contrast, participants characterized as \emph{Not Attentive} consistently reported higher workload scores. This outcome can be attributed to their need to exert additional effort to recall and process the presented concepts. This heightened mental demand, coupled with a higher level of frustration, reflects the challenges faced by individuals in this category when responding to questions associated with the viewed stimuli.
\end{itemize}
These findings highlight the crucial role of attentiveness and active engagement in shaping users' task experiences and perceived cognitive demands.

\begin{figure}[htbp]
    \centering
    \begin{subfigure}[b]{0.45\textwidth}
        \includegraphics[scale = 0.3]{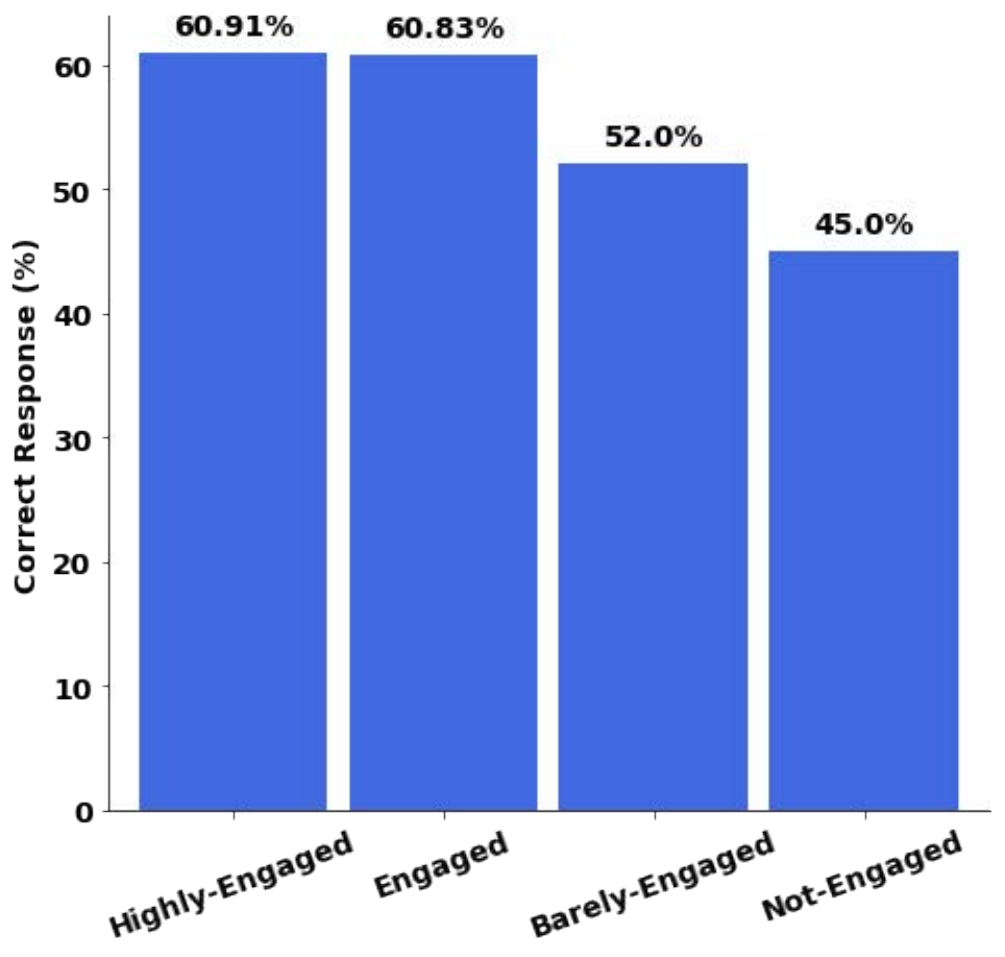}
        \caption{\% of correct responses of engagement classes.}
        \label{fig:engagement_correct_response}
    \end{subfigure}
    \hfill
    \begin{subfigure}[b]{0.45\textwidth}
        \includegraphics[scale = 0.3]{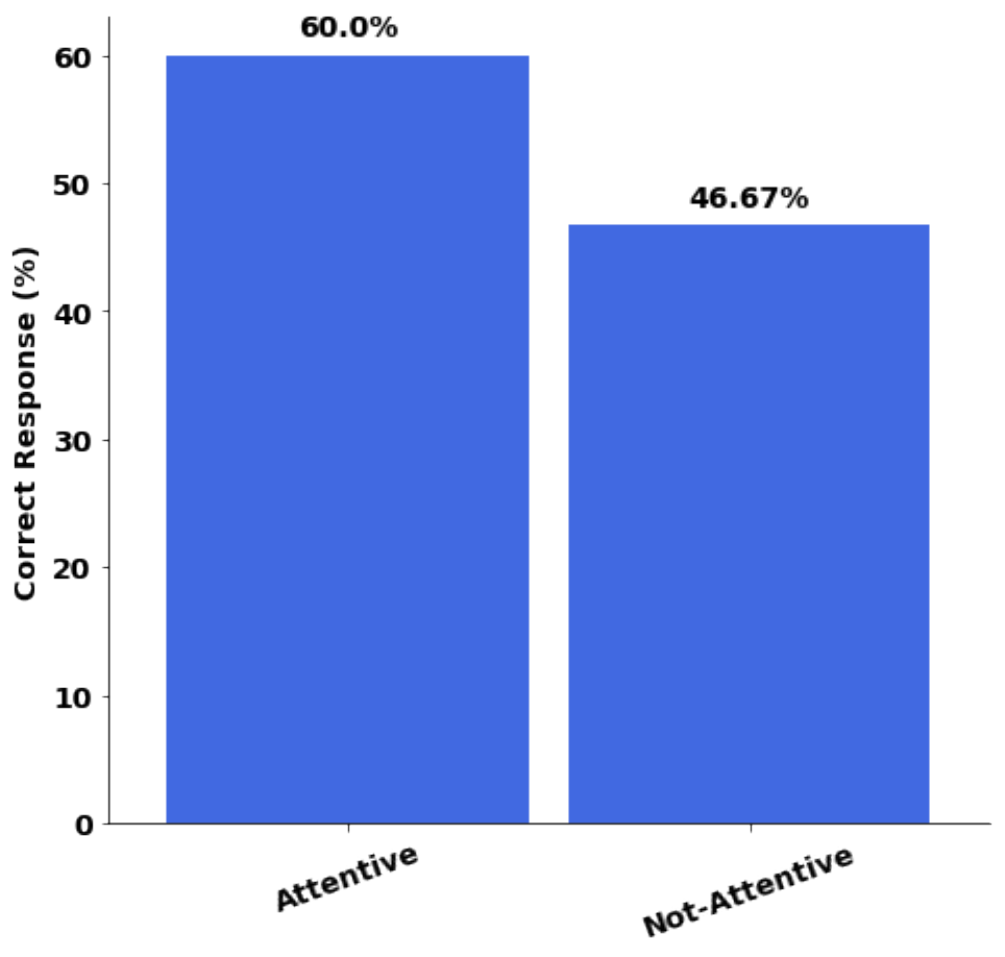}
        \caption{\% of correct responses of attention classes.}
        \label{fig:attention_correct_response}
    \end{subfigure}
    \caption{In (a) \emph{Highly Engaged} and \emph{Engaged} classes performed comparatively better than \emph{Barely Engaged} and \emph{Not Engaged} classes, whereas in (b) \emph{Attentive} class performed  comparatively better than \emph{Not Attentive} class. For details, refer to section~\ref{section:correct_responses}.}
    \label{fig:Engagement and Attention states vs Task Performance.}
\end{figure}
\vspace{-0.8cm}
\subsection{Engagement \& Attention Percentage of Correct Responses}
\label{section:correct_responses}
We assessed participants' self-labels for engagement and attention and analyzed their performance in video-based questionnaires. Each questionnaire contained two moderately difficult questions related to the stimulus content. By calculating the mode of self-labels, we assigned overall engagement and attention labels to participants. We then computed the percentage of correct responses for each label category. Figures ~\ref{fig:engagement_correct_response} and ~\ref{fig:attention_correct_response} illustrate the relationship between the percentage of correct responses and engagement and attention labels. We found that the \emph{Highly Engaged}, \emph{Engaged}, and \emph{Attentive} categories gave the highest percentage of correct responses. As engagement and attention levels decrease, the percentage of correct responses also declines. This suggests a direct relationship between task performance and the level of engagement or attention.

\section{Discussion}

In this study, we investigate the relationship between engagement and attention using the \emph{DREAMS} dataset. Our experimental results indicate that attention may be a more definitive indicator of engagement than the reverse. The underlying cause of this observation can be attributed to the fact that for Engagement, self-labels were assigned based on participants' visible reactions rather than their cognitive involvement. Conversely, attention labels were assigned based on the participants' immediate thoughts, whether related to the stimuli or not. Thus, while an individual may display visible signs of engagement their mind might be elsewhere, as illustrated by the fourth user in the attention sample frames in Figure~\ref{fig:frames}. On the other hand, attentive individuals tend to unintentionally exhibit positive signs of engagement. When these visible reactions data is fed into our model, it more readily learns that attention can lead to engagement, whereas the reverse relationship is more complex. This complexity arises because a visually engaged individual may or may not be attentive, complicating the model's ability to predict attention based on visual data.

Understanding the interplay between engagement and attention is important for optimizing user interaction with multimedia content. Our findings suggests that attention is an important factor to be considered in educational settings. The educational content should be designed to maintain the user's attention, which can improve learning outcomes. Whereas for recreational content, allowing users the freedom to engage without strict attention is beneficial, as it helps them enjoy the content and relax mentally. 

\section{Conclusion and Future Work}
We present the \emph{DREAMS} dataset, self-labelled with \emph{Engagement} and \emph{Attention} mind states. We perform experiments in single-task, transfer learning, and multi-task setups.
The results demonstrate the richness and complexity of the data and suggests that attention could be considered a clearer or more informative signal of engagement compared to the reverse relationship. Additionally, We analyzed the cognitive workload experienced by the participants of the experimental study and found that the workload was directly correlated to the complexity of the content of the video stimulus. We also observed that Higher Engagement and Attention states resulted in lower workload and vice versa. Furthermore, we analyzed the performance of the subjects belonging to various engagement and attention states and found that the percentage of correct responses was directly related to higher engagement and attentiveness. A limitation of our study is the narrow range of stimuli, primarily focusing on educational and humorous videos. This may not fully represent the broader spectrum of real-world scenarios. In the dataset extension, we will
introduce a diverse set of stimuli, including topics outside of computing, and invite participants from non-technical
backgrounds. In this work we explored attention and engagement in supervised manner. It will be intriguing to uncover this relationship in self-supervised settings. 

%
%

\begin{thebibliography}{8}
\bibitem{10.1109/FG.2018.00019}
T.~Baltrusaitis, A.~Zadeh, Y.~C. Lim, and L.-P. Morency.
\newblock Openface 2.0: Facial behavior analysis toolkit.
\newblock In {\em 2018 13th IEEE International Conference on Automatic Face \&
  Gesture Recognition (FG 2018)}, page 59–66. IEEE Press, 2018.

\bibitem{booth2017toward}
B.~M. Booth, A.~M. Ali, S.~S. Narayanan, I.~Bennett, and A.~A. Farag.
\newblock Toward active and unobtrusive engagement assessment of distance
  learners.
\newblock In {\em 2017 Seventh International Conference on Affective Computing
  and Intelligent Interaction (ACII)}, pages 470--476. IEEE, 2017.

\bibitem{10.1145/1518701.1518705}
G.~Buscher, E.~Cutrell, and M.~R. Morris.
\newblock What do you see when you're surfing? using eye tracking to predict
  salient regions of web pages.
\newblock In {\em Proceedings of the SIGCHI Conference on Human Factors in
  Computing Systems}, CHI '09, page 21–30, New York, NY, USA, 2009.
  Association for Computing Machinery.

\bibitem{cai2022marlin}
Z.~Cai, S.~Ghosh, K.~Stefanov, A.~Dhall, J.~Cai, H.~Rezatofighi, R.~Haffari,
  and M.~Hayat.
\newblock Marlin: Masked autoencoder for facial video representation learning.
\newblock In {\em Proceedings of the IEEE/CVF Conference on Computer Vision and
  Pattern Recognition (CVPR)}, pages 1493--1504, June 2023.

\bibitem{copur2022engagement}
O.~Copur, M.~Nak{\i}p, S.~Scardapane, and J.~Slowack.
\newblock Engagement detection with multi-task training in e-learning
  environments.
\newblock In {\em Image Analysis and Processing--ICIAP 2022: 21st International
  Conference, Lecce, Italy, May 23--27, 2022, Proceedings, Part III}, pages
  411--422. Springer, 2022.

\bibitem{10.1145/3242969.3264993}
A.~Dhall, A.~Kaur, R.~Goecke, and T.~Gedeon.
\newblock Emotiw 2018: Audio-video, student engagement and group-level affect
  prediction.
\newblock In {\em Proceedings of the 20th ACM International Conference on
  Multimodal Interaction}, ICMI '18, page 653–656, New York, NY, USA, 2018.
  Association for Computing Machinery.

\bibitem{10.1016/j.ijhcs.2012.01.004}
S.~D'Mello, A.~Olney, C.~Williams, and P.~Hays.
\newblock Gaze tutor: A gaze-reactive intelligent tutoring system.
\newblock {\em Int. J. Hum.-Comput. Stud.}, 70(5):377–398, may 2012.

\bibitem{10.1504/IJLT.2009.028805}
S.~K. D'Mello, S.~D. Craig, and A.~C. Graesser.
\newblock Multimethod assessment of affective experience and expression during
  deep learning.
\newblock {\em Int. J. Learn. Technol.}, 4(3/4):165–187, oct 2009.

\bibitem{grafsgaard2013automatically}
J.~Grafsgaard, J.~B. Wiggins, K.~E. Boyer, E.~N. Wiebe, and J.~Lester.
\newblock Automatically recognizing facial expression: Predicting engagement
  and frustration.
\newblock In {\em Educational data mining 2013}, 2013.

\bibitem{gupta2016daisee}
A.~Gupta, A.~D'Cunha, K.~Awasthi, and V.~Balasubramanian.
\newblock Daisee: Towards user engagement recognition in the wild.
\newblock {\em arXiv preprint arXiv:1609.01885}, 2016.

\bibitem{hart1988development}
S.~G. Hart and L.~E. Staveland.
\newblock Development of nasa-tlx (task load index): Results of empirical and
  theoretical research.
\newblock In {\em Advances in psychology}, volume~52, pages 139--183. Elsevier,
  1988.

\bibitem{heath2007we}
R.~Heath.
\newblock How do we predict advertising attention and engagement.
\newblock {\em School of Management University of Bath Working Paper}, 9, 2007.

\bibitem{10.1145/3411764.3445269}
S.~Hutt, K.~Krasich, J.~R.~Brockmole, and S.~K.~D'Mello.
\newblock Breaking out of the lab: Mitigating mind wandering with gaze-based
  attention-aware technology in classrooms.
\newblock In {\em Proceedings of the 2021 CHI Conference on Human Factors in
  Computing Systems}, CHI '21, New York, NY, USA, 2021. Association for
  Computing Machinery.

\bibitem{kaur2018prediction}
A.~Kaur, A.~Mustafa, L.~Mehta, and A.~Dhall.
\newblock Prediction and localization of student engagement in the wild.
\newblock In {\em 2018 Digital Image Computing: Techniques and Applications
  (DICTA)}, pages 1--8. IEEE, 2018.

\bibitem{lee2022predicting}
T.~Lee, D.~Kim, S.~Park, D.~Kim, and S.-J. Lee.
\newblock Predicting mind-wandering with facial videos in online lectures.
\newblock In {\em Proceedings of the IEEE/CVF Conference on Computer Vision and
  Pattern Recognition}, pages 2104--2113, 2022.

\bibitem{leiker2016relationship}
A.~M. Leiker, M.~Miller, L.~Brewer, M.~Nelson, M.~Siow, and K.~Lohse.
\newblock The relationship between engagement and neurophysiological measures
  of attention in motion-controlled video games: a randomized controlled trial.
\newblock {\em JMIR serious games}, 4(1):e5460, 2016.

\bibitem{10.1145/3577190.3614164}
M.~Singh, X.~Hoque, D.~Zeng, Y.~Wang, K.~Ikeda, and A.~Dhall.
\newblock Do i have your attention: A large scale engagement prediction dataset
  and baselines.
\newblock In {\em Proceedings of the 25th International Conference on
  Multimodal Interaction}, ICMI '23, page 174–182, New York, NY, USA, 2023.
  Association for Computing Machinery.

\bibitem{stappen2022estimation}
L.~Stappen, A.~Baird, M.~Lienhart, A.~B{\"a}tz, and B.~Schuller.
\newblock An estimation of online video user engagement from features of
  time-and value-continuous, dimensional emotions.
\newblock {\em Frontiers in Computer Science}, 4:37, 2022.

\bibitem{thomas2018predicting}
C.~Thomas, N.~Nair, and D.~B. Jayagopi.
\newblock Predicting engagement intensity in the wild using temporal
  convolutional network.
\newblock In {\em Proceedings of the 20th ACM International Conference on
  Multimodal Interaction}, pages 604--610, 2018.

\bibitem{10.5555/3295222.3295349}
A.~Vaswani, N.~Shazeer, N.~Parmar, J.~Uszkoreit, L.~Jones, A.~N. Gomez,
  L.~Kaiser, and I.~Polosukhin.
\newblock Attention is all you need.
\newblock In {\em Proceedings of the 31st International Conference on Neural
  Information Processing Systems}, NIPS'17, page 6000–6010, Red Hook, NY,
  USA, 2017. Curran Associates Inc.

\bibitem{whitehill2014faces}
J.~Whitehill, Z.~Serpell, Y.-C. Lin, A.~Foster, and J.~R. Movellan.
\newblock The faces of engagement: Automatic recognition of student
  engagementfrom facial expressions.
\newblock {\em IEEE Transactions on Affective Computing}, 5(1):86--98, 2014.

\bibitem{klinger1978modes}
E.~Klinger.
\newblock Modes of normal conscious flow.. the stream of consciousness, 1978.

\bibitem{matthews2010task}
G.~Matthews, J.~S. Warm, L.~E. Reinerman, L.~K. Langheim, and D.~J. Saxby.
\newblock Task engagement, attention, and executive control.
\newblock {\em Handbook of individual differences in cognition: Attention,
  memory, and executive control}, pages 205--230, 2010.
  
\bibitem{o2010development}
H.~L. O'Brien and E.~G. Toms.
\newblock The development and evaluation of a survey to measure user
  engagement.
\newblock {\em Journal of the American Society for Information Science and
  Technology}, 61(1):50--69, 2010.

\bibitem{reeve2011agency}
J.~Reeve and C.-M. Tseng.
\newblock Agency as a fourth aspect of students’ engagement during learning
  activities.
\newblock {\em Contemporary educational psychology}, 36(4):257--267, 2011.

\bibitem{fredricks2012measurement}
J.~A. Fredricks and W.~McColskey.
\newblock The measurement of student engagement: A comparative analysis of
  various methods and student self-report instruments.
\newblock In {\em Handbook of research on student engagement}, pages 763--782.
  Springer, 2012.


\bibitem{skinner2009motivational}
E.~A. Skinner, T.~A. Kindermann, and C.~J. Furrer.
\newblock A motivational perspective on engagement and disaffection:
  Conceptualization and assessment of children's behavioral and emotional
  participation in academic activities in the classroom.
\newblock {\em Educational and psychological measurement}, 69(3):493--525,
  2009.


\bibitem{fredricks2004school}
J.~A. Fredricks, P.~C. Blumenfeld, and A.~H. Paris.
\newblock School engagement: Potential of the concept, state of the evidence.
\newblock {\em Review of educational research}, 74(1):59--109, 2004.

\end{thebibliography}
%

\end{document}